# On times and shadows: the observational analemma

**Alejandro Gangui**
IAFE/Conicet and Universidad de Buenos Aires, Argentina

**Cecilia Lastra**
Instituto de Investigaciones CEFIEC, Universidad de Buenos Aires, Argentina

**Fernando Karaseur**
Instituto de Investigaciones CEFIEC, Universidad de Buenos Aires, Argentina

*The observation that the shadows of objects change during the course of the day and also for a fixed time during a year led curious minds to realize that the Sun could be used as a timekeeper. However, the daily motion of the Sun has some subtleties, for example, with regards to the precise time at which it crosses the meridian near noon. When the Sun is on the meridian, a clock is used to ascertain this time and a vertical stick determines the angle the Sun is above the horizon. These two measurements lead to the construction of a diagram (called an analemma) as an extremely useful resource for the teaching of astronomy. In this paper we report on the construction of this diagram from roughly weekly observations during more than a year.*

**Introduction**

Since early times, astronomers and makers of sundials have had the concept of a "mean Sun". They imagined a fictitious Sun that would always cross the celestial meridian (which is the arc joining both celestial poles through the observer's zenith) at intervals of exactly 24 hours. This may seem odd to many of us today, as we are all acquainted with the fact that one day -namely, the time it takes the Sun to cross the meridian twice- is in fact 24 hours and there is no need to invent any new Sun. However, our last statement assumes that the real Sun moves at a uniform rate along an orbit parallel to the celestial equator, covering 360° in 24h. And this is not true.

The "true Sun", which corresponds to the real Sun we observe in the sky, behaves in a more complicated way [1]. It turns out that during some weeks of the year the real day lasts a bit less than 24h, and the opposite occurs at other periods of the year. As we may imagine, if several days in a row have the property of being shorter or longer than 24h, the cumulative effect will end up being noticeable. We may synchronize our wristwatch tomorrow at noon with the passage of the Sun across the meridian and, just a few weeks later, mean (or clock) and true (or solar) noon times might be several minutes apart. The sum of the variations of the true Sun with respect to the mean Sun is encapsulated in what we now know as the *equation of time*.

In brief, for each day of the year, the equation of time (ET, hereafter) has one particular value and, in our sign convention, we define it as ET = *true time - mean time*. True time is given by sundials, while mean time is that given by our clocks and is an abstraction arising from corrections due to longitude and legalities (as we will see in the next section). If ET is positive, it means the clock is ahead of the sundial, that is, the (true) Sun runs slow and crosses the meridian later, or, another way of saying it, the sundial is behind of mean time (measured by our clocks). To offer one example, for the particular location in the city of Buenos Aires, where we performed our observations, on January 5, true noon (12h 59m) occurs 5 min after mean noon (12h 54m, which is, by the way, constant all year long). Hence, the resulting value for the equation of time for January 5 is
ET = +5 min.

On the other hand, the diurnal arc followed by the Sun across the sky changes with the seasons. Therefore, the shadows of objects observed at the same time every day will also change during the course of the year. These diurnal arcs are always parallel to the celestial equator (as they are the visible effect of our planet's rotation) and their angular separation with respect to it is called the *declination of the Sun* (δ). At the March equinox this declination is zero. It then increases and reaches the maximum value of +23°26' at the June solstice (which is the angular tilt of the Earth's axis), after which it decreases, passes through zero again at the September equinox and attains its minimum value -23°26' at the December solstice.

In summary, for each day of the year we have one particular value for ET (telling us, say, the difference between true noon and mean noon) and one particular value for δ (showing us, for example, the angular height of the Sun at true noon). So, for each day we have two quantities and we can plot them on an x-y graph, with the x-coordinate being ET and the y-axis being δ. The resulting plot is called an analemma (Fig.1).

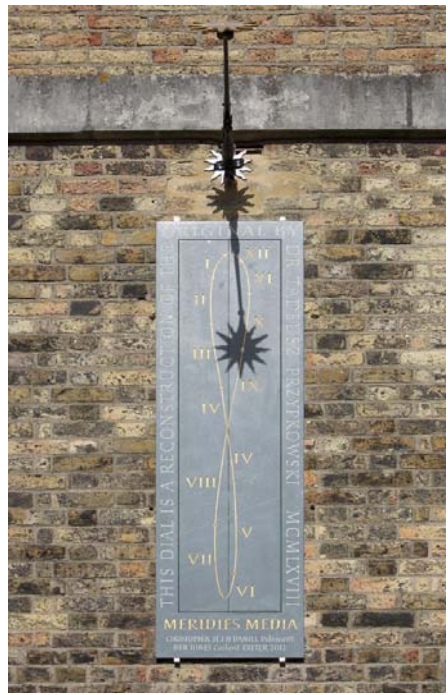

Fig. 1. A vertical mean time noon-mark (*méridienne*) at the Royal Observatory at Greenwich. This kind of sundial shows not only the true solar noon (vertical line) but also the mean noon (that is, true noon with the ET correction). When sunlight passing through the eyelet falls upon the analemma curve, local mean noon is easily read as a function of the date (Photograph source: Wikipedia; author: Christopher St J. H. Daniel).

**Diurnal astronomy during a whole year**

Beginning a couple of days before the 2015 March equinox and for more than a year, in collaboration with colleagues from a dozen South American cities, we carried out a joint shadow-observing activity with the aim of constructing the analemma. The main idea was proposed by a colleague of ours, Professor Camino, and was to share our experiences and data, and emphasize aspects that were characteristic of each location (like times of solar noon and length of the shadows) and others which were common to all cities, such as the final odd-looking elongated figure-eight-shaped graph (the analemma) each of us could finally visualize.

We all used 1-meter vertical gnomons (which are just vertical sticks projecting shadows), as it was suggested in Ref. 2. Employing the gnomon, we first drew the local meridian line on the floor, for example, using the equal shadow method [1] (the procedure detailed in [3] could also be used). Having ready the meridian line, approximately once a week we measured the angular height of the Sun at *true solar noon*, Θ(date), namely when the shadow of the gnomon crosses the meridian (as it was done in Ref. 3, but see also Ref. 4) and took note of the clock time of that moment, T(date). For simplicity, we measured the angular values using trigonometry [3]: Θ(date) = *arctan* (1 meter / length of the 1-meter gnomon shadow) . This is all we needed.

Our group made all the observations from a single site in the city of Buenos Aires, Argentina, located at coordinates (latitude, longitude) = (34°36'40.9''S, 58°29'46.3''W). Therefore, the angular height of the Sun at true solar noon during the equinoxes is Θ(equinox) = 55°23' = 90° - 34°37', where we have approximated the latitude by 34°37' (no more precision is needed) [3].

From this, for any date in which we performed our measurements (at true solar noon), we readily obtained:

δ(date) = Θ(equinox) - Θ(date) = 55°23' - Θ(date).

This is enough for the vertical spatial axis of the analemma.

Now, let us consider the horizontal temporal axis of the graph, which corresponds to the equation of time. As we mentioned, our observation site was located at longitude coordinate λ = 58°29'46.3''W = 58.4962°W. We know the Sun covers 15° in longitude in 1 hour. Therefore, the angle 58.4962° separating us from the Greenwich meridian will be covered in approximately $t_\lambda$ = 3 hours and 54 minutes. Our country fixes the official meridian to be 45° to the west of Greenwich (we belong to GMT-3 time zone). Thus, our observation site is shifted some 54 min to the west with respect to the official time meridian. This means that when the Sun crosses the local meridian line and a sundial marks solar noon, the clock time is actually 12 h 54 min (in this example we take ET = 0).

In other words, as we always make our observations at true solar noon, for any date, the equation of time can formally be computed as:

ET(date) = T(date) - (12h + Δλ),

where Δλ = $t_\lambda$ - [Time Zone] represents a fixed time value for each location. For our particular location in the city of Buenos Aires we then have: Δλ = 3 h 54 min - 3 h = 54 min.

In summary, the plot of the analemma we present in Fig. 2 has the Sun's declination δ in its vertical axis and, in its horizontal axis, the values obtained from the equation of time ET(date) = T(date) - (12 h + 54 m), where each declination and T value was measured roughly once a week.

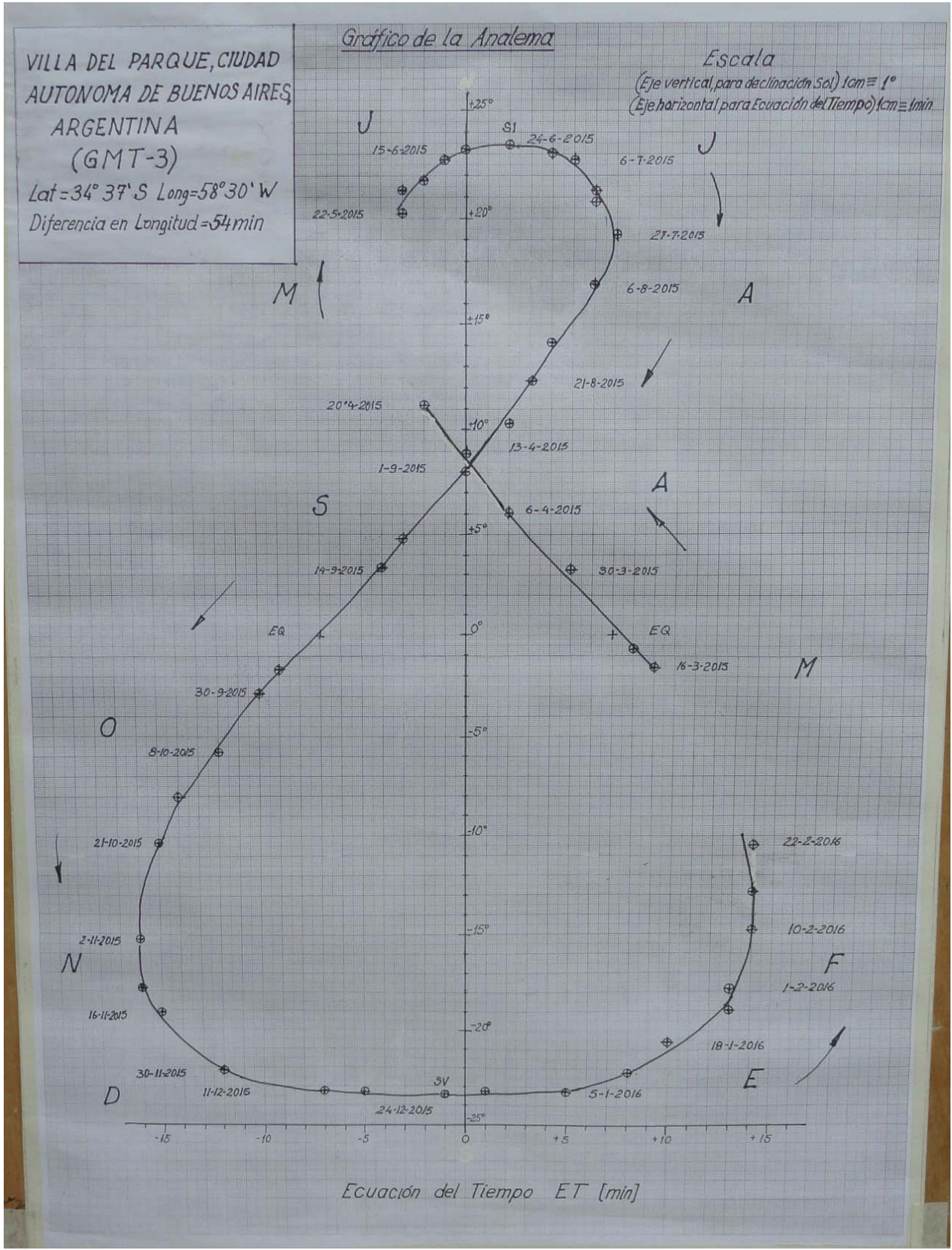

Fig. 2. The observational analemma in progress, one month before completing a full year of observations (note the gap in the period February-March). Also, some problems spoiled our data gathering during May, showing another gap that was completed in the second year of observations (see Fig. 3). The image is courtesy of Néstor Vinet.

The diagram in Fig. 2 was plotted by hand on graph paper by a colleague of ours, Professor Vinet, and shows the actual data points, together with the dates and appropriate month's labels. An approximate curve interpolating the data points (the analemma) is also shown.

However, sometimes the scales on the diagram are chosen so that equal distances represent equal angles in both directions in the sky. As the Earth rotates, relative to the Sun, at a mean speed of one degree every four minutes (or 15° in 1 hour), we can plot the data again in such a way that four minutes of time (in the horizontal axis of the equation of time) are represented by the same distance as one degree in the declination in the vertical axis.

The resulting new diagram (Fig. 3), although not as detailed as the previous one, might be more familiar to photographers who are used to superimposing pictures of the Sun and surrounding landscape taken at the same clock time of the day, from the same location, and a few days apart, for a whole year.

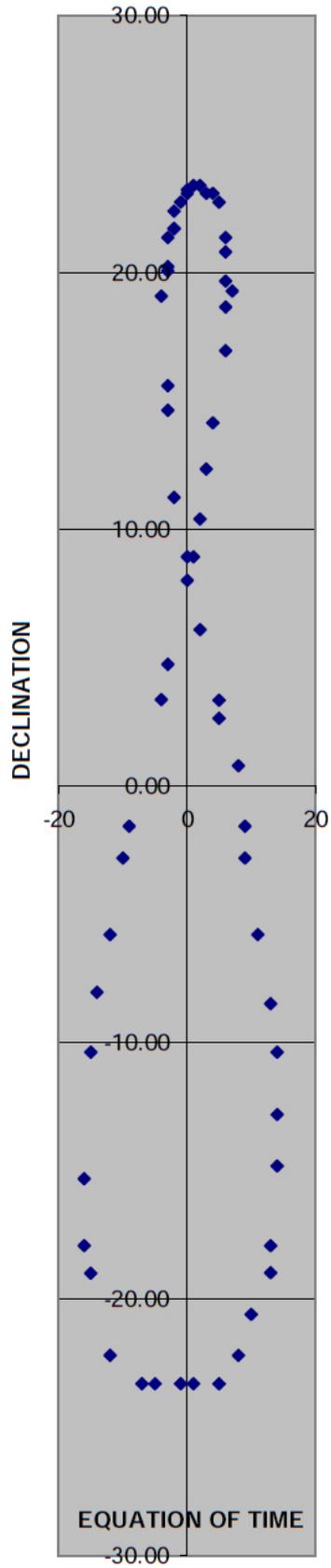

Fig. 3. The final analemma after more than 16 months of observations, from March 16, 2015, until July 28, 2016. Declination of the Sun is given in degrees, while the equation of time is given in minutes. The scale in the horizontal axis was chosen so that 4 min in ET are equivalent to 1° in declination in the vertical axis.

**Final words**

Students usually know quite well that, as described by Kepler's laws, the orbit of Earth around the Sun is not an exact circumference, and that our planet's rotation axis is not perpendicular to the plane of its orbit. This implies the subtle variation of the apparent velocity of our star when projected in the local sky and also the continuous change of the Sun's position in the sky at a fixed clock time. As we saw in this paper, both these effects can be detected by direct observations, and one way to organize the data thus obtained is already at hand for astronomy teachers, and is called the analemma.

Analemmas, those elongated figures-of-eight placed on globes of the Earth, mostly in the eastern Pacific Ocean, and on some sundials, are far from being ubiquitous in our culture. However, they offer a graphic display of one of the lessons an astronomy student might be keen to learn: the difference between clock time and Sun time. As we saw, the analemma also shows us the apparent change of the diurnal arc followed by the Sun during different seasons of the year, a clear consequence of the variation of the Sun's declination arising from the tilt of Earth's axis.

Considered an extremely useful resource for astronomy education, in this paper we reported on how to produce a diagram of the analemma from roughly weekly observations during more than a year, just employing a vertical gnomon and a clock. The activity we described here can easily be replicated elsewhere. Most interestingly, doing joint observations in collaboration with schools located in far away cities will help students recognize that although the actual measurements (namely, the angular heights of the Sun and true solar noon times) will turn out to be different between observers, the final plot of the analemma will be, apart from minor observational uncertainties, the very same for all. In fact, the analemma arises from a property of the Sun-Earth system and is therefore independent of a particular observation site.

**Acknowledgements**

We thank our colleagues from the project "observational construction of the analemma", especially Professors N. Camino and N. Vinet, for sharing with us their observations and graphs. We also appreciate the valuable input of the four referees who have contributed to substantially improving our manuscript. The first author acknowledges support from CONICET (The National Scientific and Technical Research Council in Argentina) and the University of Buenos Aires.

Alejandro Gangui *is staff researcher at the Institute for Astronomy and Space Physics (IAFE) and is professor of physics at the University of Buenos Aires (UBA).* gangui@iafe.uba.ar

Cecilia Lastra *is a Physics Teacher graduated from CEFIEC, University of Buenos Aires.*

Fernando Karaseur *is currently finishing his degree as a Physics Teacher at CEFIEC, University of Buenos Aires.*